\documentclass[aps,nofootinbib,11pt,preprintnumbers]{revtex4}
\usepackage{amsfonts,amssymb,amsmath}
\usepackage{bm}		% bold math
\usepackage{epsfig}
\def\d{\partial}
\def\n{{\bf n}}

\def\x{{\bf x}}
\def\q{{\bf q}}
\def\vs{v_{\rm s}}
\def\dx{{\bf dx}}
\def\al{\bm\alpha}
\def\grad{{\bm\nabla}}

\def\A{{\scriptscriptstyle A}}
\def\B{{\scriptscriptstyle B}}
\def\M{{\scriptscriptstyle M}}
\def\N{{\scriptscriptstyle N}}
\def\T{{\bf T}}

\def\be{\begin{equation}}
\def\ee{\end{equation}}

\begin{document}

\preprint{PUPT-2281}
\preprint{INT PUB 08-45}

\title{Holographic model of superfluidity}
\author{C.~P.~Herzog}
\affiliation
  {Department of Physics, Princeton University, Princeton, NJ 08544, USA}
\author{P.~K.~Kovtun}
\affiliation
  {Department of Physics and Astronomy, University of Victoria,
  Victoria, BC, V8P 5C2, Canada}
\author{D.~T.~Son}
\affiliation
  {Institute for Nuclear Theory, University of Washington, Seattle, WA
  98195-1550, USA}

\begin{abstract}
\noindent
We study a holographic model of a relativistic quantum system
with a global $U(1)$ symmetry, at non-zero temperature and density.
When the temperature falls below a critical value,
we find a second-order superfluid
phase transition with mean-field critical exponents.
In the symmetry-broken phase, we determine the speed of second
sound as a function of temperature.
As the velocity of the superfluid component relative to the
normal component increases, the superfluid transition
goes through a tricritical point and becomes first-order.
\end{abstract}
\maketitle

\section{Introduction}
\noindent
Recently, gauge/gravity
duality~\cite{Maldacena:1997re,Gubser:1998bc,Witten:1998qj} has been used
to model strongly-interacting systems in terms of a gravity dual. The most
important case is the strongly-interacting quark gluon plasma created at
RHIC.  While a systematic treatment is still lacking, the ${\cal N}=4$
super-Yang-Mills plasma has been used instead, with some success.  Most
notably, the viscosity/entropy density ratio, universal among all theories
with a gravity dual~\cite{Kovtun:2004de,Buchel:2003tz,Policastro:2001yc},
seems to be not too far away from the value extracted from experimental
data~\cite{Romatschke:2007mq,Luzum:2008cw}.

The long-distance behavior of a black hole horizon is captured by the same
hydrodynamic equation governing the evolution of the plasma.  As a
result, from the gravity equations one can reconstruct the hydrodynamic
equations, including transport coefficients.  One can go beyond 
leading order in the gravity equations and capture second-order
corrections, in the process discovering extra terms typically ignored in
almost all implementations of the Israel-Stewart
formalism~\cite{Baier:2007ix,Bhattacharyya:2008jc}.  The approach has been
extended to R-charged black holes, dual to fluids with chemical
potentials~\cite{Erdmenger:2008rm,Banerjee:2008th}.

In this work, we use black holes to study relativistic superfluids.
Our work is a continuation of ref.\ \cite{HHH}, where a holographic
model of a superfluid was constructed.  Here, we
investigate the behavior of the superfluid when a superfluid current
flows through the system.  It is known that in nonrelativistic
superfluids there is a critical superfluid velocity, above which the
superfluid phase does not exist.  We discover exactly the same
phenomenon in the relativistic superfluids at low temperatures: there
is a first-order phase transition between the superfluid and the
normal phase as one changes the superfluid current.  The phase
transition becomes second order at higher temperatures.

We will also be discussing the superfluid hydrodynamics of such systems,
in particular, the speed of second sound.
A short discussion of hydrodynamics can be found in \cite{Son}.
Hydrodynamics of a system with spontaneously broken symmetries
is different from hydrodynamics of normal liquids because
the system contains long-range modes (Goldstone bosons)
which must be included in the hydrodynamic equations.

\section{Thermodynamics}
\label{sec:thermo}
\noindent
Let us first review the thermodynamics of quantum systems with
a spontaneous $U(1)$ symmetry breaking.

For simplicity consider a system with one global $U(1)$ symmetry,
such as complex $\phi^4$ theory.
We define the free energy $F = - T \ln Z$ where
\be
Z = {\rm tr} \, e^{-\beta H} \ ,
\ee
and $H$ is the Hamiltonian. 
While we eventually take the thermodynamic limit $V\to\infty$,
for now it is more convenient to think of the system as living inside
a $d-1$ dimensional torus $\T^{d-1}$ where the perimeters 
of the $d-1$ circles are $L_i$.  

When the $U(1)$ symmetry is not spontaneously broken,
the free energy $F$ depends only on the temperature, and, 
if we do not take the thermodynamic limit, the $L_i$: $F=F(T, L_i)$.
When the symmetry is spontaneously broken, the system
contains a massless (pseudo) scalar field $\varphi$, which
is compact: $\varphi\sim\varphi+2\pi$.
In the example of the complex $\phi^4$ theory, $\varphi$
is just the phase of $\phi$.
If one thinks about $\varphi$ as the elementary field,
periodicity in $\varphi$ space means that we are free to
impose boundary conditions such that
$\varphi|_{x_i{=}L_i} = \varphi|_{x_i{=}0} + 2\pi n_i$.
Thus there is a set of partition functions $Z[\n]$,
where we integrate over all fields with boundary conditions
specified by $\n$.  The usual ground state corresponds to $n_i=0$.

As we shall see later, $Z[\n]$ does not exist in the strict
thermodynamic sense.  Physically, if one sets $n_i\neq0$, the system
will relax over time to a state with no winding.  However, the time
scale for this process may be long, and the winding state is
practically stable.  We assume stability.

Define $q_i \equiv \d_i \varphi$,
an analogue of the topological charge density
whose integral can distinguish various boundary conditions:
\begin{equation}
 \oint_{C_i} \q \cdot \dx = \int_0^{L_i} q_i \, dx^i  = 
 \varphi|_{x_i=L_i} - \varphi|_{x_i=0} = 2 \pi n_i \ ,
\end{equation}
where $C_i$ is the $i$'th spatial circle of the $\T^{d-1}$.
We can therefore impose the boundary conditions on $\varphi$
by introducing Lagrange multipliers $\lambda_i$ which pick out the
required values of $n_i$:
\begin{equation}
  Z[\n] = \int_{\{\n\}}\!\! D\phi \; e^{-S_E[\phi]}
        = \int\!\! D\phi\, d\lambda \; e^{-S_E[\phi]}\; 
          \exp \Biggl( i \sum_j \lambda_j \biggl( \oint_{C_j} \q \cdot \dx 
  - 2\pi n_j \biggr) \Biggr) \ .
\end{equation}
We are taking advantage of the fact that the Euclidean path integral 
on a compact time circle
of perimeter $\beta = 1/T$ yields the thermal partition function 
at a temperature $T$.
Having compactified the time direction, it is natural to ask what 
boundary conditions should
be applied for $\varphi$ in the time like direction.  The answer 
to this question involves 
introducing a chemical potential, and we will return to it shortly.

We can write the field $\varphi$ as
$\varphi(x)=\alpha_i x^i + \delta\varphi(x)$ where $\alpha_i$
is constant, and $\delta\varphi(x)$ is periodic.
The delta function in the path integral for $Z[\n]$ will pick out
$\alpha_i=2\pi n_i/L_i$, so knowing $\alpha_i$ is equivalent to
knowing the boundary conditions for $\varphi$.
In the thermodynamic limit, $F$ depends on the $L_i$ only
through an overall multiplicative volume factor $V$ and 
$\alpha_i$ becomes a continuous variable
which transforms as a vector under rotations.
At zero temperature, the energy (for given $\n$) is always minimized
for $\varphi=\alpha_i x^i$ because that's when the $(\nabla\varphi)^2$
term is smallest.
So at zero temperature, the ground state can be uniquely
characterized by the $\alpha_i$.
We assume now that the thermal state can be uniquely characterized
by $\alpha_i$ and $T$, and define the partition function as
\begin{equation}
  Z[{\al}]
  = \int\!\! D\phi\, d\lambda \; e^{-S_E[\phi]}\; 
    \exp\Biggl( i \sum_j \lambda_j \oint_{C_j} (\grad \varphi - \al) \cdot \dx 
  \Biggr) \,.
\label{eq:Z}
\end{equation}
Note that $\al$ is the equilbrium value of $(\grad\!\varphi)$.
In nonrelativistic superfluids $\grad\!\varphi$ is proportional to the
superfluid velocity.  We shall use the same terminology and call 
$\grad\!\varphi$ the superfluid velocity.
$Z[\al]$ describes a stationary state
with constant, non-zero superfluid velocity.
The partition function $Z$ is a scalar, and therefore can only
depend on $\al^2$ in the thermodynamic limit.

Note that the state described by $Z[\al]$ is in fact metastable,
rather than stable.  This is because $\varphi$, as the phase of a
complex field $\phi$, is ill-defined when $\phi=0$; hence the winding
numbers $n_i$ are not topological and can change with time.  The process
of ``unwinding'' can be visualized as follows.  Imagine a state with
only one of the $n_i$ equal to 1, and other winding numbers equal 0.
The state can be thought of as containing a
$(d{-}2)$ dimensional domain wall, across which $\varphi$ changes by $2\pi$.
However, in $d>2$ dimensions, there is no topological 
conservation law which would ensure the stability of the
domain wall, and the domain wall will decay through
hole nucleation.
In terms of superfluid hydrodynamics in 3+1 dimensions,
nucleating a hole in the two-dimensional domain wall
corresponds to producing a vortex loop.
Production of vortex loops violates Landau's criterion
for superfluidity \cite{Feynman}, and renders the state with
non-zero $\al$ metastable.

In a theory with large $N$, like the one we will be considering, the
rate of vortex loop production is exponentially suppressed, as both
the energy of a vortex loop, and the action of the configuration
describing vortex nucleations, are proportional to $N$.
Therefore, provided that the volume and time scales are not too large,
one can treat states with a nonzero superfluid velocity as thermal
equilibrium states.

We now would like to introduce a chemical potential $\mu$ and work in the
grand canonical ensemble.  
We define the potential function $\Omega \equiv F- \mu Q =  - PV
=- T \ln {\mathcal Z} $ where
\be
{\mathcal Z} = {\rm tr} \, e^{-\beta(H-\mu Q)}
\label{eq:calZ}
\ee
and $Q$ is the conserved charge corresponding to the $U(1)$ symmetry.
To introduce $\mu$ at the level of the path integral,
it is useful to gauge the $U(1)$ symmetry by
coupling the system to a non-dynamical gauge field $A_\mu$:
\be
{\mathcal Z}[A] = e^{W[A]} = \int D \phi \, e^{-S_E[\phi, A] } \ .
\label{eq:ZA}
\ee
We are temporarily ignoring the dependence of ${\mathcal Z}$ on the
$\al$.
The one point function of the U(1) current is then generated by $W[A]$:
\be
\frac{\delta W[A]}{\delta A_\mu(x)} = \langle J^\mu(x) \rangle \ .
\label{onepoint}
\ee
In a system where the external field strength is zero $F^{\mu\nu}=0$, we may
pick a gauge in which $A_\mu$ is constant and it makes sense, 
given (\ref{onepoint}),
 to interpret $A_0 = \mu$ as the chemical potential.

Gauge transformations which send $A_\mu\to A_\mu+\d_\mu\lambda$ 
have a nontrivial
effect on the compact scalar: $\varphi\to\varphi+\lambda$.
In the presence of the background gauge field, 
the pressure $P$ can depend only on gauge invariant quantities such as
$D_\mu \varphi = \d_\mu \varphi - A_\mu$.
The grand canonical partition function thus becomes
\be
 {\mathcal Z}[A,{\al}]
  = \int\!\! D\phi\, d\lambda \; e^{-S_E[\phi, A]}\; 
    \exp\Biggl( i \sum_j \lambda_j \oint_{C_j} (\grad \varphi -{\bf A}- \al) 
  \cdot \dx \Biggr) \,.
\label{eq:calZpartition}
\ee
Like $Z[\al]$, by rotational symmetry 
${\mathcal Z}[\al]$ can only depend on $\al^2$ in the thermodynamic limit.
We are assuming that the pressure $P=TV^{-1}\ln{\cal Z}$ 
depends on three parameters,
$P=P(T,\mu,\frac12 (D_i \varphi)^2)$.

We can now choose a gauge in which $\varphi{=}0$, so that
$(D_i\varphi)^2=A_i^2$.
The chemical potential $\mu$ can be interpreted as the zero
component of the gauge field, and we have $P=P(T,A_0,A_i^2)$.
At zero temperature, the system is Lorentz invariant,
provided one treats $A_0$ and $A_i$ as spurion fields that transform 
as a four-vector under the Lorentz group.
As a result, 
the pressure can only depend on $A_\mu^2=A_i^2{-}A_0^2$.
At finite temperature, the pressure depends on $A_0$ and
$A_i^2$ separately, which we can write as
$P=P(T,A_0,A_\mu^2)$.  
In other words, gauging the symmetry allows us to trade
the dependence of pressure on the condensate phase
for the dependence of pressure on the background gauge
field.  In a general gauge, $P=P(T,-D_0\varphi,(D_\mu\varphi)^2)$.
Denoting $\chi\equiv\frac12(D_\mu\varphi)^2$, we have the pressure
as a function of three thermodynamic variables:
$P=P(T,\mu,\chi)$.  From it one can define the conjugate variables,
\begin{equation}\label{eq:dP}
  dP = s\,dT + n\,d\mu - f^2 d\chi\,.
\end{equation}
Notice that $s$, $n$, and $f^2$ are functions of $T$, $\mu$, and $\chi$.
If the third term were absent from the Eq.~(\ref{eq:dP}), then 
$s$ would be the entropy density, and $n$ the charge density.  We will
give the interpretation of $s$, $n$ and $f^2$ in the next section.

\section{Hydrodynamics}
\label{sec:hydro}
\noindent
Let us set the external gauge field $A_\mu$ to 0.
The equations of ideal relativistic superfluid hydrodynamics are known.
We will use the form written in~\cite{Son}.  In this formulation, the
degrees of freedom are $T$, $\mu$, $\varphi$, and a unit four-vector $u^\mu$
satisfying $\eta_{\mu\nu}u^\mu u^\nu=-1$ (recall that we use the 
mostly plus convention for the flat metric tensor $\eta_{\mu\nu}$).  
We will later identify
$u^\mu$ with the velocity of the normal component.
The equations consist of the conservations of the energy-momentum
tensor $T^{\mu\nu}$ and of the $U(1)$ symmetry current $j^\mu$,
\begin{eqnarray}
&& \d_\mu T^{\mu\nu}=0\,,
\label{eq:Tmunu-conservation}\\
&& \d_\mu j^\mu=0\,.
\label{eq:jmu-conservation}
\end{eqnarray}
and a ``Josephson equation'' describing time evolution of $\varphi$, 
\begin{equation}
  u^\mu \d_\mu\varphi + \mu =0\,.
\label{eq:mu-definition}
\end{equation}
The stress-energy tensor and the current are expressed in terms of the
hydrodynamic variables through
\begin{eqnarray}
&& T^{\mu\nu} = (\epsilon+P)u^\mu u^\nu + P \eta^{\mu\nu} 
                + f^2\, \d^\mu\varphi\; \d^\nu\varphi\,,
  \label{eq:Tmunu-definition}\\
&& j^\mu = n u^\mu + f^2 \d^\mu\varphi\label{eq:jmu-definition}\,.
\end{eqnarray}
where the energy density $\epsilon$ is defined by $\epsilon+P=Ts+n\mu$.
Thus we have $(d{+}2)$ hydrodynamic equations
(\ref{eq:Tmunu-conservation}), (\ref{eq:jmu-conservation}),
(\ref{eq:mu-definition}) for $(d{+}2)$ variables
$T$, $\mu$, $\varphi$ and $u^\mu$. % (recall that $u^\mu u_\mu=-1$).
One can derive from 
Eqs.~(\ref{eq:Tmunu-conservation})--(\ref{eq:jmu-definition})
\begin{equation}
  \d_\mu(su^\mu) =0.
  \label{eq:s-conservation}
\end{equation}
One can interpret this equation as the equation of entropy
conservation (entropy is conserved since we are working at the level
of ideal, nonviscous hydrodynamics).  Thus $s$ is the entropy density
and $u^\mu$ is the velocity of entropy flow.  In the two-fluid model
only the normal component carries entropy; therefore $u^\mu$ is
interpreted as the normal velocity.  The two contributions to the
current in Eq.~(\ref{eq:jmu-definition}) can be interpreted as the 
normal and superfluid currents.  Therefore $n$ is the normal
density, and $f^2$ is the analogue of the pion decay constant. (The
equivalent of the superfluid density would be $f^2\mu$.)

Let us look at small fluctuations about an equilibrium state
at fixed temperature, chemical potential and zero 
normal and superfluid velocities,
i.e.\ we write $T=T_0+T'$, $\mu=\mu_0+\mu'$, 
 $u^\mu=(1,{v^i}')$, $\xi_i\equiv\d_i\varphi=\xi_i'$,
where $T'$, $\mu'$, $v^i$, and $\xi'$ are small.  In terms of $\xi$ 
the variation of pressure is
$dP = s\,dT + (n{+}\mu f^2)d\mu - f^2 \xi d\xi$,
where $\xi=|\vec\xi|$.
Further, let us assume that pressure
is a smooth function of $\xi^2$ at small $\xi$,
so that $\d P/\d\xi=0$ in equilibrium. 
The linearized hydrodynamic equations become
(omiting subscript ``0'' on equilibrium quantities)
\begin{subequations}
\label{eq:linear-sf-hydro}
\begin{eqnarray}
&& \frac{\d^2P}{\d T\d\mu}\, \d_t T' +
   \frac{\d^2P}{\d\mu^2}\, \d_t \mu' +
   f^2\, \d_i\xi_i' + n\, \d_i v_i' = 0\,,\\
&& \left(\mu\frac{\d^2P}{\d T\d\mu}+T\frac{\d^2P}{\d T^2}\right) \d_t T' +
   \left(\mu\frac{\d^2P}{\d\mu^2}+T\frac{\d^2P}{\d T\d\mu}\right) \d_t\mu'+
   w\,\d_i v_i' + f^2\mu\; \d_i\xi_i' = 0\,,\\
&& w\, \d_t v_i' + f^2 \mu\, \d_t\xi_i' +
   s\, \d_i T' + (n{+}\mu f^2)\,\d_i\mu' = 0\,,\\
&& \d_t\xi_i'+\d_i\mu'=0\,.
\end{eqnarray}
\end{subequations}
In this system of equations, pressure is taken as $P=P(T,\mu,\xi)$,
and $w=\epsilon+P$ is the density of enthalpy.
The first equation is the linearized current conservation
equation $\d_\mu j^\mu=0$.
The second equation is the linearized energy conservation
$\d_\mu T^{\mu 0}=0$.
The third equation is the linearized momentum conservation,
$\d_\mu T^{\mu i}=0$.
Finally, the fourth equation says that $\mu$ and $\xi_i$ are
not independent because $\mu=-\d_t\varphi$ while $\xi_i = \d_i\varphi$.

An interesting feature of the linearized hydrodynamic equations
(\ref{eq:linear-sf-hydro}) is that they admit propagating mode
solutions even if one ignores the fluctuations of the energy-momentum
tensor, i.e. if one ignores the (normal) sound fluctuations.
If one were to ignore the condition $\d_\mu T^{\mu\nu}=0$,
together with fluctuations of temperature $T'$
and velocity of the normal component $v_i'$, then the
system (\ref{eq:linear-sf-hydro}) becomes
\begin{subequations}
\begin{eqnarray}
&&  \frac{\d^2P}{\d\mu^2}\, \d_t \mu' + f^2\, \d_i\xi_i' =0\,,\\
&&  \d_t\xi_i'+\d_i\mu'=0\,.
\end{eqnarray}
\label{eq:linearprobe}
\end{subequations}
Fourier transforming all variables as $e^{-i\omega t+i k\cdot x}$,
we find a propagating mode with frequency
\begin{equation}
\label{eq:second-sound-leftover}
    \omega^2 = v_2^2 k^2\,, \quad\quad
    v_2^2    = \frac{f^2}{\left(\frac{\d^2 P}{\d\mu^2}\right)}
             = -\frac{\left(\frac{\d^2 P}{\d\xi^2}\right)_{\!T,\mu}}
                     {\left(\frac{\d^2 P}{\d\mu^2}\right)_{\!T,\xi}}
             \;>0 \,.
\end{equation}
When expressing $f^2$ in terms of $(\d^2 P/\d\xi^2)$, we
have assumed that $P=O(\xi^2)$ at small $\xi$.
The thermodynamic derivatives in Eq.~(\ref{eq:second-sound-leftover})
are to be evaluated at $\xi=0$.
The propagating mode (\ref{eq:second-sound-leftover})
is the leftover of the second sound
in superfluids which survives even if one ignores the
$\d_\mu T^{\mu\nu}=0$ part of the hydrodynamic equations.
One expects
that this mode is captured by
the dual gravitational description which ignores the backreaction
of the gauge fields on the metric.

It is not difficult to find the eigenmodes of the full system
(\ref{eq:linear-sf-hydro}).
Taking all variables proportional to $e^{-i\omega t+i k\cdot x}$,
one finds a fourth order equation for frequency,
\begin{equation}
\label{eq:w4}
  a\, \omega^4 - b\, k^2 \omega^2 + c\, k^4 = 0 
\end{equation}
where the coefficients $a$, $b$ and $c$ 
are independent of $k$.
When $b^2>4ac$, the equation for $\omega^2$
has two real positive roots $\omega^2=\vs^2 k^2$
and $\omega^2=v_2^2 k^2$.
The first solution is the normal sound,
and the second solution is the second sound.
In terms of thermodynamic derivatives,
the coefficients $a$, $b$ and $c$ are given by
\begin{eqnarray}
&& a = Tw\left[\left(\frac{\d^2 P}{\d T^2}\right)
               \left(\frac{\d^2 P}{\d \mu^2}\right)
              -\left(\frac{\d^2 P}{\d T \d\mu}\right)^2\right]\,,\\
&& b = \left(\frac{\d^2 P}{\d T^2}\right)\! T (n^2{+}wf^2) +
       \left(\frac{\d^2 P}{\d \mu^2}\right) \! T s^2 -
      2\left(\frac{\d^2 P}{\d T\d\mu}\right)\! Ts\,n \,, \\
&& c = f^2 T s^2\,.
\end{eqnarray}
In particular, $a$, $b$, and $c$ are positive.
Let us now look at simple examples.
In the non-superfluid phase $f^2=0$, and therefore
$\vs^2=b/a$, while the second sound is absent, $v_2^2=0$.
Even in the non-superfluid phase,
the speed of the normal sound looks like a complicated
expression in terms of the derivatives of pressure $P(T,\mu)$.
The expression for $\vs^2$ simplifies if instead of $P(T,\mu)$
we work with $P(s,n)$.
Indeed, the coefficient $a$ is proportional to the Jacobian of
the transformation from the $(s,n)$ variables
to the $(T,\mu)$ variables.
In terms of $P(s,n)$, the speed of sound in the normal phase becomes
\begin{equation}
\label{eq:vs-sn}
  \vs^2 = \frac{n}{w} \left(\frac{\d P}{\d n}\right)_{\!\!s} + 
          \frac{s}{w} \left(\frac{\d P}{\d s}\right)_{\!\!n}\,.
\end{equation}
Now that we have $\vs^2$ expressed in terms of $P(s,n)$
without reference to the chemical potential, we can
evaluate the thermodynamic derivatives in the canonical
ensemble instead of the grand canonical.
In the canonical ensemble, the total charge (number of particles) $N=nV$
is fixed, and therefore $dn/n=-dV/V$.
It will be convenient to go from the $(s,n)$ to the $(S,\epsilon)$ variables.
For the total entropy $S$ we have
$
  {dS}/{V} = {d(sV)}/{V} = ds - {s}\, dn/n\,,
$
while on the other hand the relation $TdS=dE+PdV$ gives
$
  {T}\,dS/V = d\epsilon - {w}\, dn/n\,.
$
This implies that the speed of sound
in the normal phase (\ref{eq:vs-sn}) can be written as
\begin{equation}
  \vs^2 = \left(\frac{\d P}{\d\epsilon}\right)_{\!S,N}\,.
\end{equation}
As another example, consider a scale-invariant theory,
in which case pressure has the form $P(T,\mu)=T^d g(T/\mu)$, where
$g(T/\mu)$ is a dimensionless function.
The speed of normal sound evaluated from Eq.~(\ref{eq:w4})
is $\vs^2=1/(d{-}1)$, in either normal or superfluid phase.
The speed of the second sound evaluated from Eq.~(\ref{eq:w4})
can be written as
\begin{equation}
\label{eq:second-sound-CFT}
  v_2^2 = {f^2}\left[\left(1+\frac{\mu n}{Ts}\right) 
                     \left(\frac{\d^2 P}{\d\mu^2} - \frac{n+\mu f^2}{s}
                     \frac{\d^2 P}{\d T\d\mu}\right)\right]^{-1}\,.
\end{equation}
We can see that in scale invariant theories,
the cartoon expression (\ref{eq:second-sound-leftover})
can be recovered in the formal ``large-entropy''
limit $Ts\gg\mu n$, together with
$s(\frac{\d^2P}{\d\mu^2})\gg(n{+}\mu f^2)(\frac{\d^2P}{\d T\d\mu})$.
In theories which do not have scale invariance, the speed of
the second sound does not have the simple form (\ref{eq:second-sound-CFT}),
but can be easily determined from Eq.~(\ref{eq:w4})
given the equation of state $P(T,\mu)$.

Note that the simple expression (\ref{eq:second-sound-leftover})
ceases to be a good approximation to the speed of the
second sound at $T\ll\mu$.
This is because at small temperatures we expect that
the system can be described as a gas of massless Goldstone bosons,
with entropy density $s\sim T^{d-1}$.
On the other hand, we expect in the low-temperature region that
$n\sim\mu^{d-1}$ which imples that the condition
of large entropy can not be satisfied at $T\ll\mu$.
At low temperatures (high densities) the speed of the
second sound has to be determined from the full equation
(\ref{eq:w4}).

\section{Dual gravity description}
On the gravity side, we study the Einstein-Maxwell theory
with a complex scalar field.
Fluctuations of the bulk metric correspond to fluctuations
of $T^{\mu\nu}$ on the boundary, while fluctuations of the
$U(1)$ gauge field correspond to fluctuations of $J^\mu$
on the boundary.
The scalar field is charged under the bulk $U(1)$, and its
background value corresponds to the condensate on the boundary.
The mass of the scalar is a free parameter.
We also need to specify the boundary conditions.
The metric will be asymptotically AdS because we want to
study the superfluid system using the AdS/CFT correspondence.
The boundary conditions for the scalar correspond to the
``normalizable'' mode because we want to describe
the system in which a charged operator has a vev.
However, the boundary conditions for the $U(1)$ gauge field
should correspond to the ``non-normalizable'' mode because
we want to describe the system in the background gauge
field, as discussed above.
In other words, we fix the value of $A_\mu$ at the
asymptotic AdS infinity.
Fixing the value of $A_0$ at infinity amounts to fixing
the chemical potential in the boundary theory, and leads
to a charged black hole in AdS.
Fixing the value of $A_i$ at infinity does not introduce
additional ``hair'' for the black hole because we expect
the new black hole solution to be only metastable,
in accord with field theory expectations.
So we want to find a (metastable) stationary solution with
the above boundary conditions, and then study small 
fluctuations around this solution, corresponding to
hydrodynamic fluctuations in the boundary theory.

In the following, we will ignore the backreaction of the
gauge and scalar fields on the metric.
It would be nice to include the backreaction,
at least perturbatively to leading order.
The action is
\begin{equation}
  S = -\int\!\! d^{d+1}\!x\; \sqrt{-g}
      \left[\frac{1}{4e^2} F_{\!\M\N}F^{\M\N} 
      + (D_{\!\M}\phi)(D^\M\phi)^* 
      + m^2 \phi \phi^* \right]\,,
      \label{gravaction}
\end{equation}
where $D_\M\phi=\d_\M\phi-iA_\M\phi$,
$F_{\M\N}=\d_\M A_\N-\d_\N A_\M$,
and capital Latin indices run from $0$ to $d$.
The equations of motion are
\begin{equation}
   \frac{1}{\sqrt{-g}}\, D_{\!\A} \big(\sqrt{-g}\, 
   g^{\A\B} D_{\!\B}\phi \big) = m^2 \phi\,,
\label{eq:phi-eqn}
\end{equation}
\begin{equation}
   \frac{1}{\sqrt{-g}}\, \d_\M 
   \big(\sqrt{-g}\, g^{\M\A} g^{\N\B} F_{\A\B}\big)
   = e^2\; g^{\N\A} J_\A\,,
\label{eq:A-eqn}
\end{equation}
where the current is 
$J_\A = i[\phi^* (D_\A\phi) - \phi (D_\A\phi)^*]$.
We write the bulk scalar as
$\phi = \frac{1}{\sqrt{2}}\rho e^{i\varphi}$, and
make a gauge transformation $A_\M\to A_\M+\d_\M\varphi$.
In the new gauge, the phase $\varphi$ disappears from
the equations of motion, and the current becomes
\begin{equation}
  J_\M = \rho^2 A_\M\,.
\end{equation}
From the Maxwell equations (\ref{eq:A-eqn}) one can see
that the background value for $\rho$ would induce
a (position-dependent) mass for the gauge field.
This is the Higgs mechanism in the bulk.
We take the $(d{+}1)$ dimensional background metric
to be of the following form:
\begin{equation}
  ds^2 = \frac{1}{z^2} \big({-}f(z)dt^2 + d\x^2 
         + \frac{dz^2}{f(z)}\big)\,.
\label{metric}
\end{equation}
The metric with $f(z)=1$ corresponds to pure AdS and
$f(z) = 1-(z/z_h)^d$ corresponds to our black hole solution.
To find the background solution for $\rho$ and $A_\M$,
we take all fields independent of $t$ and $\x$.
The $z$-component of the Maxwell equations (\ref{eq:A-eqn})
now gives $\rho^2 A_z=0$ which means we can choose $A_z=0$.
The equations of motion become
\begin{equation} \label{rhoeq}
  z^{d-1} \d_z \left[ \frac{f}{z^{d-1}}\rho'\right]
  = \left(A_i^2 - \frac{A_t^2}{f} + \frac{m^2}{z^2} \right)\rho\,,
\end{equation}
\begin{equation}
  z^{d-3} \d_z\! \left[\frac{1}{z^{d-3}}\, A_t'\right] 
  = \frac{e^2}{z^2 f}\, \rho^2 A_t\,,
  \label{Ateq}
\end{equation}
\begin{equation}
  z^{d-3} \d_z\! \left[\frac{f}{z^{d-3}}\, A_i'\right] 
  = \frac{e^2}{z^2}\, \rho^2 A_i\,.
  \label{Aieq}
\end{equation}
In the limit when $A_i=0$ they reduce to the coupled
($A_t$, $\rho$) system of equations studied recently 
in \cite{HHH}. 
Note that we have a coupled system of non-linear ODEs ---
it would be interesting to see if the solutions may exhibit
chaotic behavior.

\subsection*{The free energy}

Up to boundary counter terms, the free energy of the field theory 
is determined by the value of the
action (\ref{gravaction}) evaluated on-shell, 
$\Omega = -T S_{\rm os} + \ldots$, where the ellipsis denotes 
boundary terms that we will presently introduce. 
 Employing the equations of 
motion, we may rewrite (\ref{gravaction}) as
\be
S_{\rm os} =  
\int d^{d}x \left[ \frac{\sqrt{-g} g^{zz}}{2} \left. 
 \left( \frac{1}{e^2} g^{\mu\nu} A_\mu A_\nu'
+ \rho \rho' \right) \right|_{z=\epsilon} +\frac{1}{2} 
 \int_\epsilon^{z_h} dz \sqrt{-g} A_\mu A^\mu  \rho^2 \right] \ .
\ee
We have included a cut-off $\epsilon$ because this on-shell action 
is naively divergent
and needs to be regularized through the addition of boundary terms.  
It is difficult to treat the general case both succinctly and clearly 
and several full treatments already exist in the literature 
(see for example \cite{Skenderisreview}).  
We proceed to regularize this action in the simple case $d=3$ and $m^2 = -2$.
Using the explicit form of the metric (\ref{metric}), 
the on-shell action reduces to
\be
S_{\rm os} = \int d^d x \left[  
\left. \frac{f}{2} \left( \frac{1}{e^2 } \eta^{\mu\nu} A_\mu A_\nu' + 
\frac{1}{z^2} \rho \rho' \right) \right|_{z=\epsilon} +
\frac{1}{2} \int_\epsilon^{z_h}  dz\,  \sqrt{-g} \,  A_\mu A^\mu
 \rho^2 \right] \ .
\ee

The near boundary behavior of the fields takes the form
\begin{eqnarray}
A_\mu &=&  ( a_\mu+ {\mathcal O}(z^2)) + z (b_\mu+ {\mathcal O}(z^2)) \ , \\
\rho &=& z (a+{\mathcal O}(z^2)) + z^2  (b+ {\mathcal O}(z^2)) \ .
\end{eqnarray}
These boundary values have various reinterpretations in the field theory.  
For the gauge field, 
$a_0 = \mu$ is the chemical potential while $a_i = -\xi_i$ is 
a superfluid velocity.
Then $b_0 \sim -n$ is proportional to the charge density while 
$b_i \sim J_i$ are charge currents.
For the scalar field, there exists an ambiguity \cite{KlebanovWitten}.  
For a scalar operator $O_1$ of conformal dimension one, 
$a = \langle O_1 \rangle$ while $b$ is a source.  For a scalar $O_2$ 
of conformal dimension two
$b = \langle O_2 \rangle$ while $a$ is a source.  

In regulating $S_{\rm os}$, we must carefully formulate 
the boundary conditions.  
For example, do we wish to keep $A_\mu$ or $A_\mu'$ fixed on the boundary?  
Keeping $A_0$ fixed corresponds to keeping the chemical potential fixed 
and thus working in the grand canonical ensemble in the field theory. 
In the spatially homogenous case where we can set $A_z=0$, varying $A_M$ 
in the bulk 
 leads to a boundary term proportional to $ A_\mu' \delta A_\mu$.  
Without an additional boundary term, we are working in an ensemble 
where $a_\mu$ is held fixed.  If we would like to work in the canonical 
ensemble, at fixed charge, it is
$b_0$ that must be held fixed at the boundary ($\delta A_0' = 0$).  To 
accommodate this change, we would need to make what amounts to 
a Legendre transform and add the boundary term
\begin{equation}
\left. \frac{1}{e^2} \int d^d x A_0 A_0' \right|_{z=\epsilon}  = -\mu Q/T \ ,
\end{equation}
to the action (\ref{gravaction}). 
We will work in an ensemble where $a_\mu$ is held fixed and thus need 
no such further boundary terms.

A similar decision needs to be made about the scalar operator. 
It is most natural to work in an ensemble where the value of 
$\langle O_i \rangle$ is fixed instead of the source for the operator.
However, we still must decide whether we want our scalar operator 
to have conformal
dimension one or two.  As the ensemble where the source for $O_1$ 
is fixed is equivalent
to the ensemble where $\langle O_2 \rangle$ is fixed, we know that 
the ensembles where
$\langle O_1 \rangle$ and $\langle O_2 \rangle$ are fixed must be 
related by a Legendre transform 
\cite{KlebanovWitten}.

Consider first the case where $\langle O_1 \rangle$ is fixed.  
The on-shell action is naively divergent,
and we must add a counter-term.  The counter-term and the $\rho \rho'$ 
term in the on-shell action combine to give
\begin{equation}
 \left.
 \left(
 \frac{1}{2z^2} \rho \rho' 
 -  \frac{1}{2z^3} \rho^2 \right) \right|_{z=\epsilon} 
 = \frac{1}{2} a b + {\mathcal O}(\epsilon) \ .
\end{equation}
  In the Legendre transformed case, where $\langle O_2 \rangle$ is fixed,
we add two counter-terms, one to ensure that we hold $\delta \rho'$ 
fixed at the boundary instead of
$\delta \rho$ and one to control the divergence:
\begin{equation}
  \left(
  \left. \frac{1}{2z^2} \rho \rho' -\frac{1}{z^2} \rho \rho' 
  + \frac{1}{2z^3} \rho^2 \right) \right|_{z=\epsilon} 
  = - \frac{1}{2} ab + {\mathcal O}(\epsilon) \ .
\end{equation}
Recalling that this regularized on-shell action is the negative of 
the free energy and assuming a spatially homogenous system so that 
we may divide out by a factor of the volume $V$, we find for the 
free energies that 
\be
\Omega_i(\mu, \xi, O_i)/V = \frac{1}{2}
\left[ -\mu n + \xi \cdot J + (-1)^i O_1O_2 - \int_0^{z_h}
A_\mu A_\nu g^{\mu\nu} \rho^2  \sqrt{-g} \, dz 
\right] \ .
\label{Omega}
\ee
Recall $\Omega = - PV$.
Having fixed $O_i$, $\mu$ and $\xi$, the conjugate boundary values 
$\epsilon_{ij} O_j$, $Q$, 
and $J_s$ are then determined through the dynamics of the gravitational theory.

Since $\Omega_1(O_1)$ and $\Omega_2(O_2) = \Omega_1(O_1) + O_1 O_2 V$ are
Legendre transforms of each other, we have 
\begin{equation}
\frac{1}{V} \frac{\partial \Omega_1}{\partial O_1} = -O_2 \; \; \; 
\mbox{and} \; \; \;
\frac{1}{V} \frac{\partial \Omega_2}{\partial O_2} = O_1 \ .
\end{equation}
Thus critical points of the free energies correspond to gravitational 
solutions where at least 
one of the two $O_i$ vanish.

\subsection*{Numerical results}

\begin{figure}[h]
\begin{center}
a) \epsfig{file=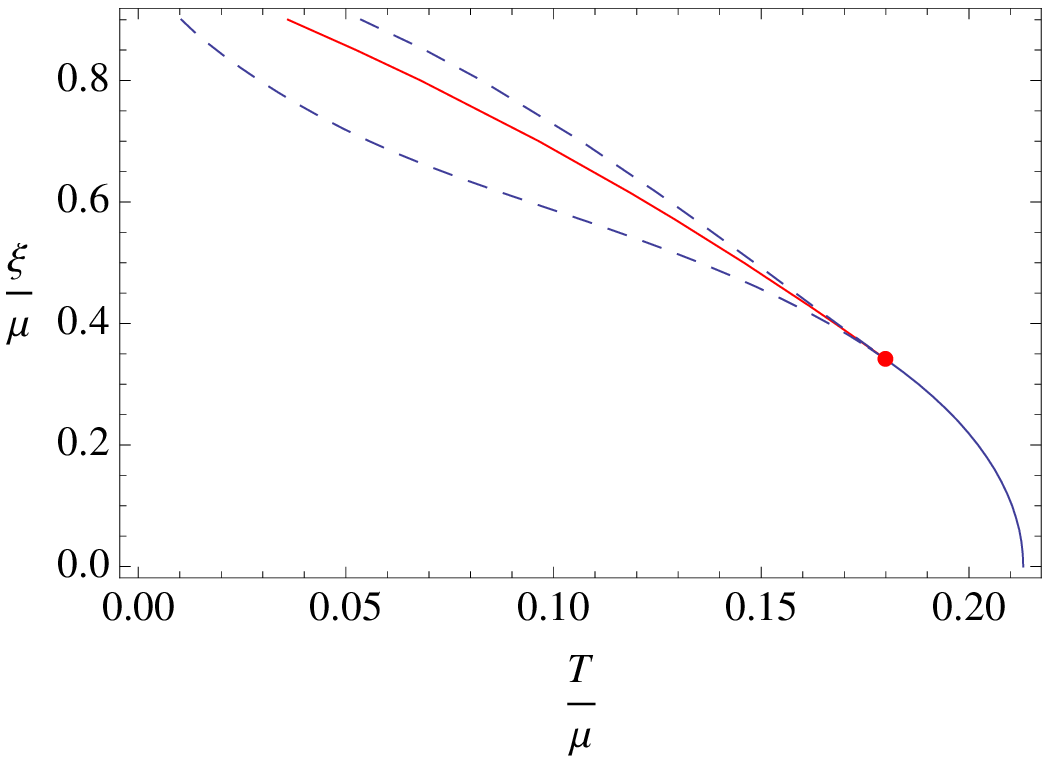,width=2.8in,angle=0,trim=0 0 0 0}%
\hspace{0.2cm}
b) \epsfig{file=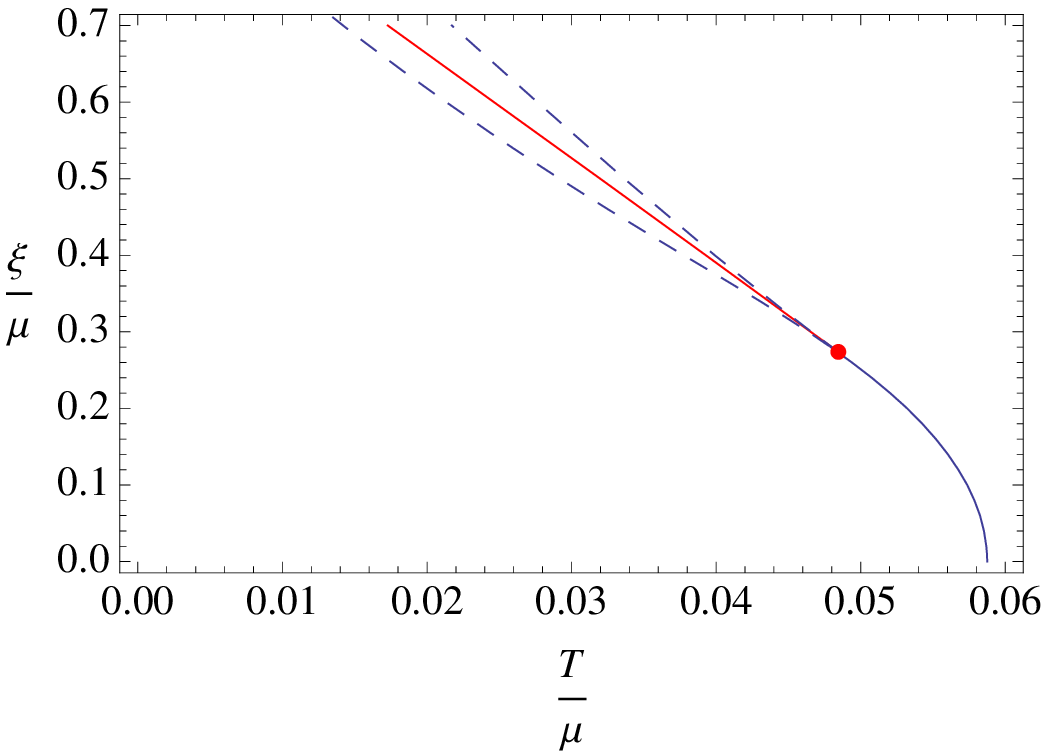,width=2.8in,angle=0,trim=0 0 0 0}%
\end{center}
\caption{The phase diagrams for the theory with a scalar with 
a) conformal dimension one
 and b) conformal dimension two.  The solid blue line indicates a 
second order phase transition while the solid red line (in between the dashed lines) 
indicates a 
first order phase transition.  The dashed blue lines are spinodal 
curves, while the red dot indicates the tricritical point.
  \label{fig:pd}}
\end{figure}

The nonlinear differential equations (\ref{rhoeq}--\ref{Aieq}) appear 
to be intractable analytically.
However, it is relatively straightforward to integrate the equations 
numerically.  The results of this section are most succinctly summarized
 by the two phase diagrams for the scalars $O_1$ and $O_2$ shown
in Figure~\ref{fig:pd}.\footnote{%
 We set $e=1$ in this section.
}

\begin{figure}[h]
\begin{center}
a) \epsfig{file=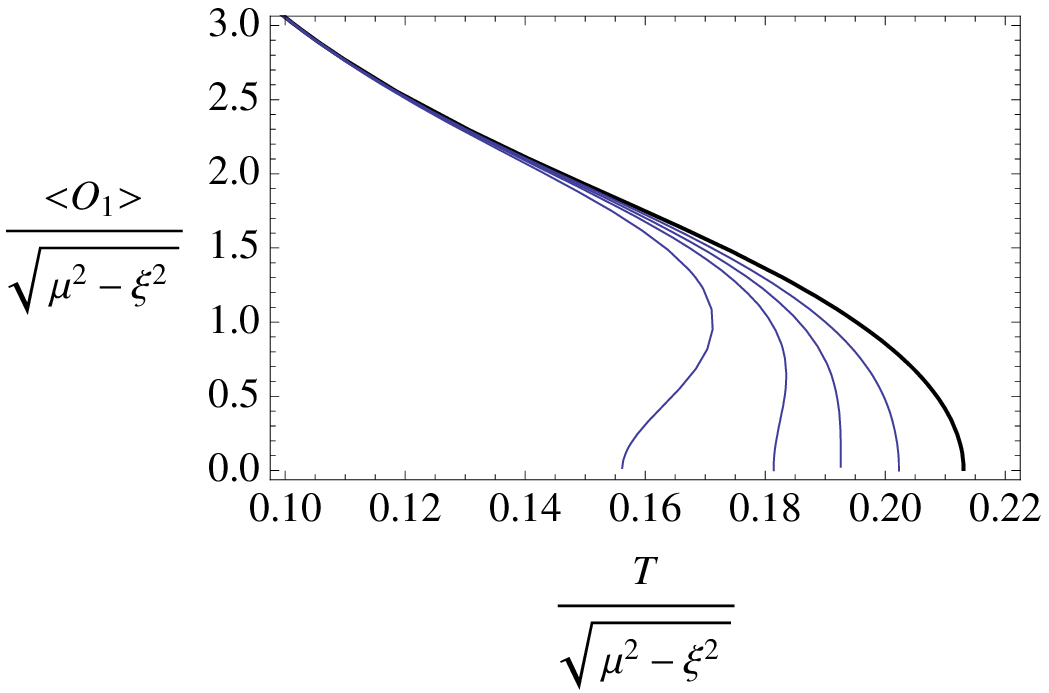,width=2.8in,angle=0,trim=0 0 0 0}%
\hspace{0.2cm}
b) \epsfig{file=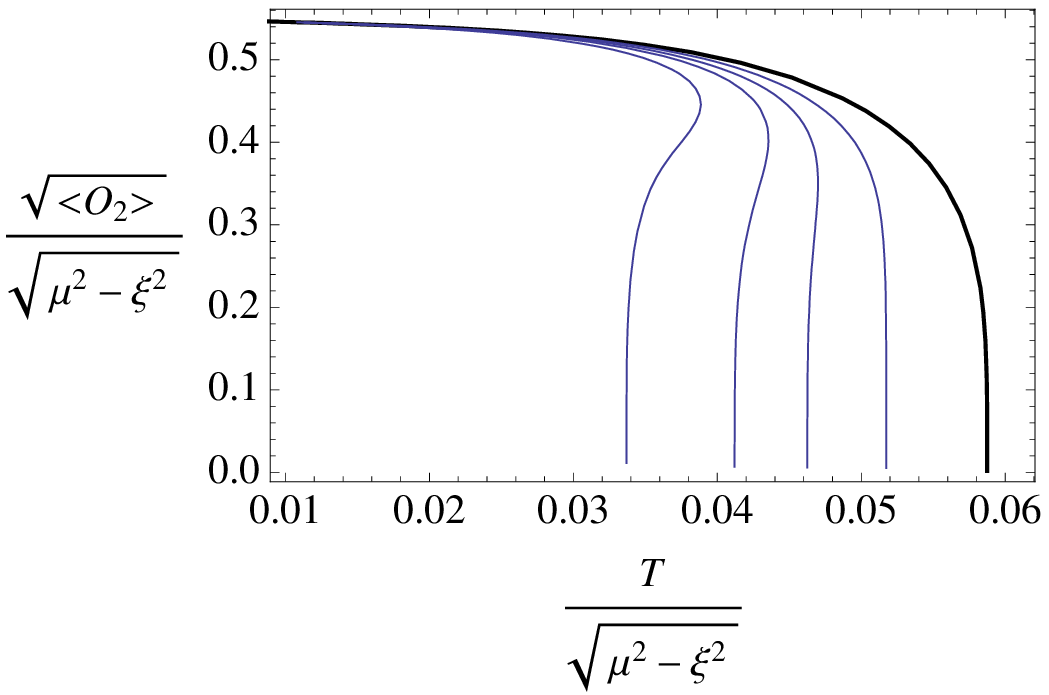,width=2.8in,angle=0,trim=0 0 0 0}%
\end{center}
\caption{The condensate as a function of
temperature for the two operators: (a) $O_1$ and (b) $O_2$.
The curves in the plots, from right to left, are for $\xi / \mu = 0$, 
$1/4$, $1/3$, $2/5$, and $1/2$.
  \label{fig:condensate}}
\end{figure}

\begin{figure}[h]
\begin{center}
a) \epsfig{file=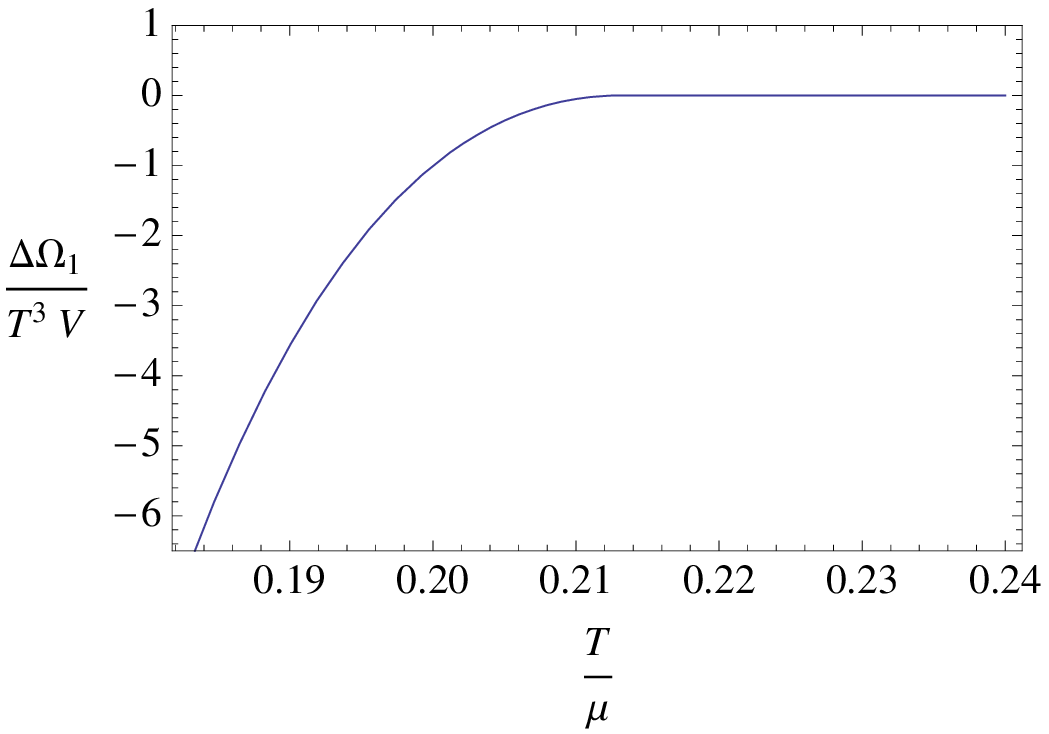,width=2.8in,angle=0,trim=0 0 0 0}%
\hspace{0.2cm}
b) \epsfig{file=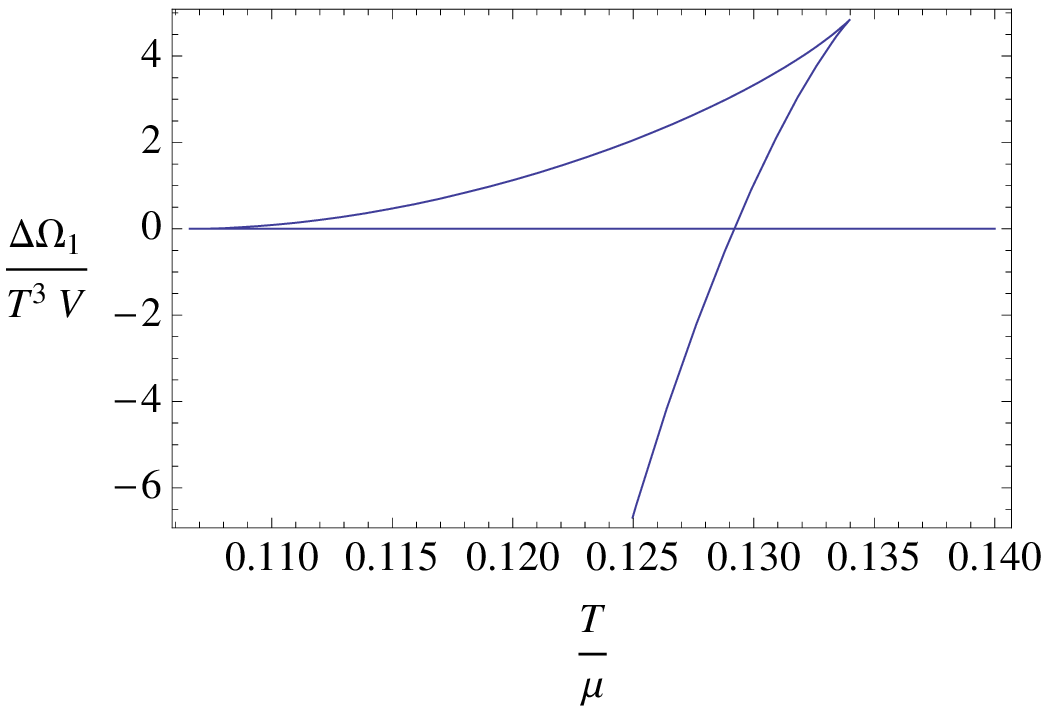,width=2.8in,angle=0,trim=0 0 0 0}%
\end{center}
\caption{The difference in free energy $\Delta \Omega_1$ between the phase 
with a scalar condensate and without one as a function of $T/\mu$:   
a) $\xi=0$ and b) $\xi/\mu = 4/7$.
  \label{fig:phasetransition}}
\end{figure}

To see where these phase diagrams come from, 
we begin by reviewing the case $\xi=0$ studied in the canonical ensemble 
in \cite{HHH} while here we choose to work at fixed $\mu$.  Given $\xi=0$, 
the third differential equation
(\ref{Aieq}) drops out.  Plots of the expectation value 
$\langle O_i \rangle$ versus temperature are shown in 
Figure \ref{fig:condensate}, as the black curves on the far right.  
At high temperature,
$\langle O_i \rangle =0$, but at  
the critical temperature $T_c$,
there is a second order phase transition where the expectation values 
become nonzero. 
For $O_1$, $T_c = 0.213 \mu$ while for $O_2$, $T_c = 0.0587 \mu$.
Near but slightly below $T_c$ 
the scalars exhibit the standard mean field
scaling with the reduced temperature
\begin{equation}
 \langle O_i  \rangle \sim (T_c - T)^{1/2} \ .
\end{equation}
The most straightforward way to see that the phase transition is 
second order is to examine a plot
of the free energy versus temperature: $\Omega_i$ is smooth at $T_c$.  
Using eq.~(\ref{Omega}), we have produced Figure \ref{fig:phasetransition}a 
for the scalar $O_1$, 
which indeed shows this smooth behavior.  We do not show a very similar 
plot for the 
second scalar $O_2$.

A more elaborate demonstration that the phase transition is second order 
comes from an investigation of
$\Omega_i$ as a function of the order parameter $O_i$ near $T_c$, recovering 
completely standard results in the Landau-Ginzburg mean field theory of 
phase transitions.  We have found numerically that 
\be
\frac{\Delta \Omega_i}{ V \mu^3} = \alpha_i(T) 
\left( \frac{\langle O_i \rangle}{|\mu|^i} \right)^2 + 
\beta_i(T) \left( \frac {\langle O_i \rangle}{|\mu|^i} \right)^4
\ee
fits the free energy curves extremely well.  Moreover, $\alpha_i$ and 
$\beta_i$ are nearly linear in 
$T$:
\begin{eqnarray}
\alpha_1 = 3.07 (T-T_c) /T_c&,& \beta_1 = 0.743 - 0.899 (T-T_c) /T_c \ ,\\
\alpha_2 = 5.13 (T-T_c) /T_c &,& \beta_2 = 1.34 - 1.45 (T-T_c) /T_c \ .
\end{eqnarray}
By definition, $\alpha_i$ passes through zero at the phase transition.

As we increase the superfluid velocity $\xi$, nothing dramatic happens 
immediately.  
The phase transition remains second order although $T_c$ decreases as 
can be seen from Figure \ref{fig:condensate}.  Because the decrease is 
due to the additional kinetic energy of the system, we expect the 
decrease to be quadratic in $\xi$, which is born out by the shape of 
the second order lines 
in the phase diagram, Figure \ref{fig:pd}.
Numerically, we find that
$T_c(\xi) \approx T_c(0) - \lambda \xi^2 / \mu$ where $\lambda 
\approx 0.27$ for $O_1$
and $\lambda \approx 0.14$ for $O_2$.

However, there exists a critical $\xi$ above which the phase transition 
becomes first order.
For $O_1$, this critical velocity is $\xi = 0.342 \mu$ while for $O_2$ 
it is $\xi = 0.274 \mu$.
From Figure \ref{fig:condensate}, it is clear that something interesting 
must happen because
the curves $\langle O_i \rangle$ become multi-valued for sufficiently 
large $\xi$.  
It is possible to see that the phase transition is first order in 
different ways.  The simplest is
to look at the free energy as a function of temperature.  
Figure \ref{fig:phasetransition}b presents
this classic swallow tail shape for $O_1$ and $\xi / \mu = 4/7$.  
At the phase transition, the free energy is continuous but not
differentiable.  The two nonanalytic points in the free energy curve 
are ``spinodal points" or 
points beyond which one
of the phases ceases to exist even as a metastable minimum of the free energy.

To demonstrate more convincingly that the phase transition becomes 
first order, we computed the free energy as a function of the
 order parameter near the putative tricritical point.  We 
found numerically that the free energy is well described by the 
sixth order polynomial
\be
\frac{\Delta \Omega_i}{ V \mu^3} = \alpha_i(T, \xi) 
\left( \frac{\langle O_i \rangle}{|\mu|^i} \right)^2 + 
\beta_i(T,\xi) \left( \frac {\langle O_i \rangle}{|\mu|^i} \right)^4
+\gamma_i(T, \xi) \left( \frac {\langle O_i \rangle}{|\mu|^i} \right)^6 \ .
\ee
At the tricritical point, $\alpha_i$ and $\beta_i$ both vanish.
Moreover, near the tricritical point, $\alpha_i$ and $\beta_i$ 
vary linearly with $T$ and
$\xi$ while $\gamma_i$ is positive and roughly constant.

\begin{figure}
\centerline{a) \includegraphics[width=2.65in]{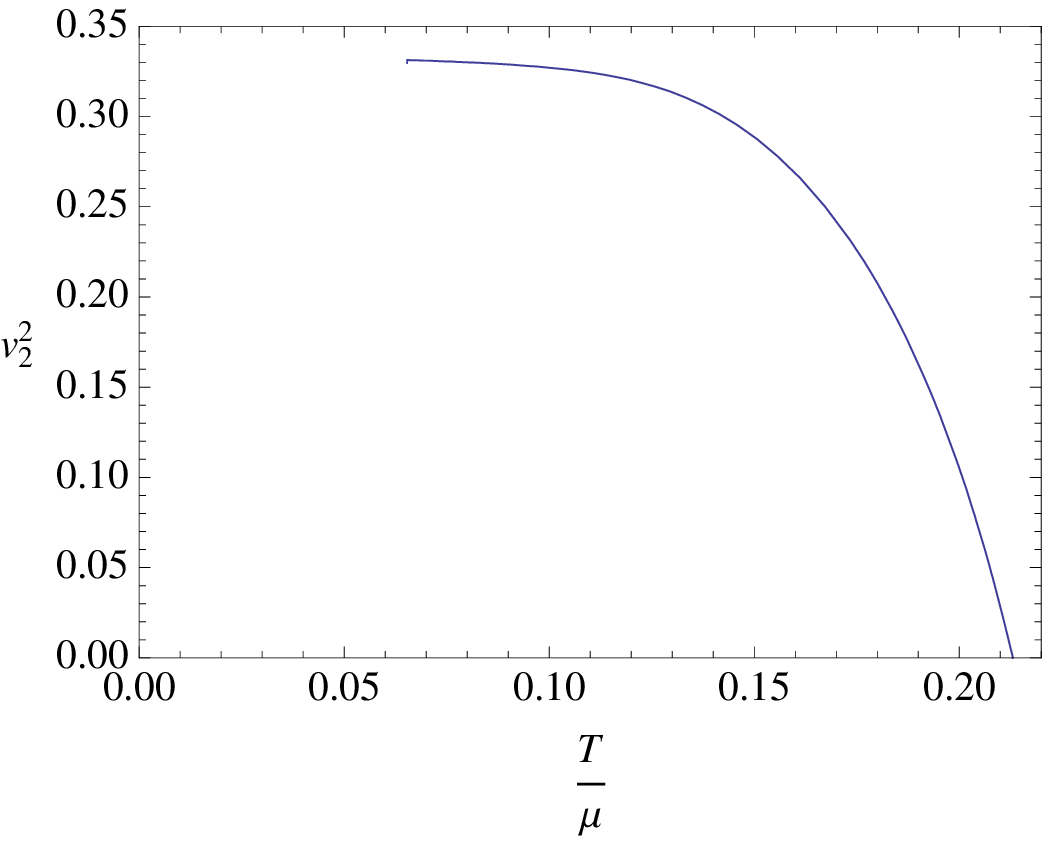}
\hskip 0.2in
b) \includegraphics[width=2.74in]{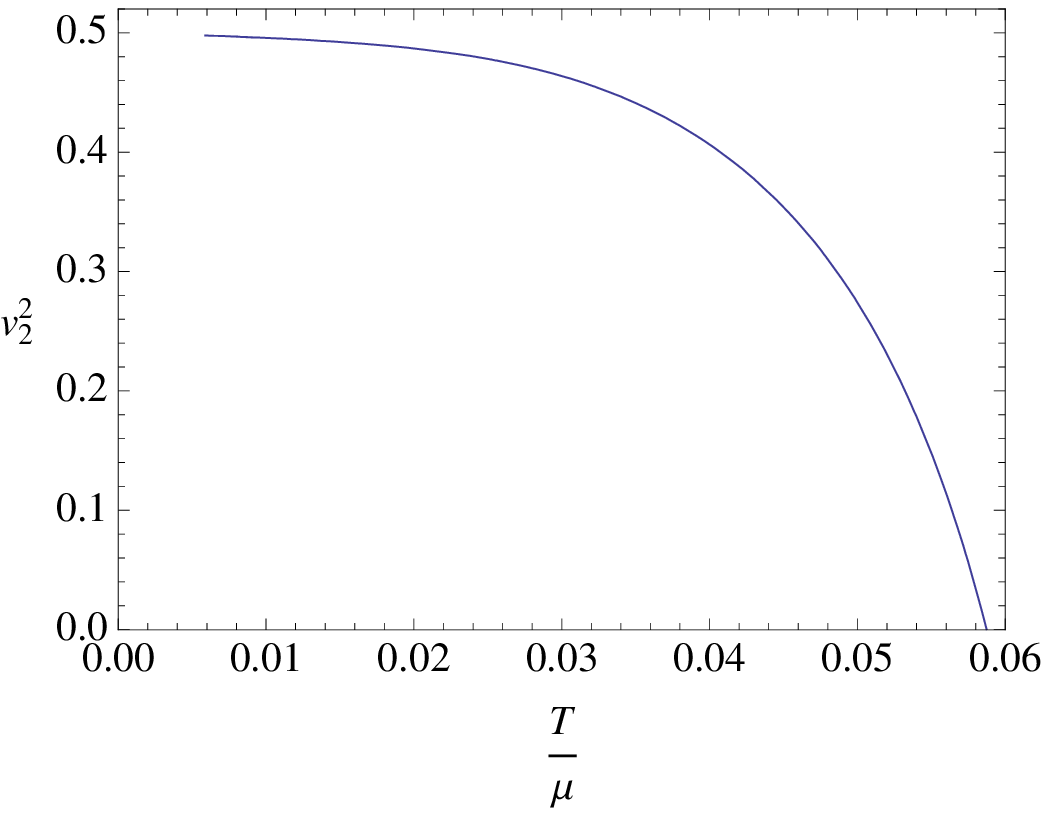}
}
\caption{
	The speed of second sound as a function of $T/\mu$,
	computed by evaluating thermodynamic derivatives in 
	Eq.~(\ref{eq:second-sound-leftover}): a) $O_1$ scalar, b) $O_2$ scalar.
	The speed of second sound vanishes as $T\to T_c$ and appears to
	approach a constant value as $T\to 0$.}
\label{fig:v2-vs-T}
\end{figure}

Before moving to a discussion of second sound, we would like to point out 
one nice feature of Fig.~\ref{fig:condensate}:  The curves approach 
each other at low temperature.  This agreement is a consequence of 
the fact that at $T=0$, the pressure $P$ can only be a function of the 
Lorentz invariant quantity $\mu^2 - \xi^2$, as discussed in 
Section \ref{sec:thermo}.

The speed of second sound as a function of temperature is shown
in Fig.~\ref{fig:v2-vs-T}.  
The plots were computed in the probe approximation using 
Eq.\ (\ref{eq:second-sound-leftover}).
We have set the superfluid velocity $\xi=0$.
The behavior close to $T_c$ is qualitatively similar to that of 
superfluid ${}^4$He \cite{oldpaper}.  As the temperature is decreased
 from $T_c$, 
the speed of second sound rises rapidly
from zero and eventually levels off.
Experimentally, it is difficult to go to very low temperature 
and remain within the hydrodynamic limit.  
The scattering length for the phonons approaches the system size.
Numerically, we have also had difficulty finding solutions at very low temperatures;
the curves in Fig.\ \ref{fig:v2-vs-T} terminate where our numerics fail.
Theoretically, however, the expectation for ${}^4$He is that second sound at 
$T \approx 0$ is a sound wave supported by a gas of phonons 
\cite{superfluidbook}.  
Thus, the speed of second sound close to 
$T=0$ should approach $v_2^2 = v_s^2/(d-1)$.   

The naive extrapolation of the curves
in Fig.\ \ref{fig:v2-vs-T} suggest that $v_2^2$ is $1/3$ in the $O_1$ case
and $1/2$ in the $O_2$ case.
The fact that the curves in Fig.~\ref{fig:v2-vs-T} level off at low temperature is caused
numerically by a similar growth in the 
susceptibility and the pion decay constant in Eq.\ (\ref{eq:second-sound-leftover}).
That we work in the probe approximation suggests our model may not be 
reliable at low temperature anyway.  
We have neglected the 
coupling between our Abelian-Higgs sector and the metric.  The 
Abelian-Higgs sector on its own does not support ordinary sound.

\section{Concluding Remarks}

We have discussed the hydrodynamics of relativistic superfluids.  
Moreover, we presented a holographic model that reproduces 
many of the familiar features of superfluids, including 
a critical superfluid velocity above which the system returns 
to its normal phase and a second order phase transition 
at zero superfluid velocity between the normal and superfluid phases.  

In many respects, our holographic model is similar 
to the Landau-Ginzburg mean field treatment of phase transitions.  
Near the phase transition, we saw that the potential function 
for the order parameter has the familiar polynomial expansion 
$V(\phi) = \alpha \phi^2 + \beta \phi^4 + \ldots$.  
In our case, however, the values of $\alpha$ and $\beta$ 
and indeed the entire structure of $V(\phi)$ 
are not set directly by the field theorist 
but are instead encoded in a nontrivial fashion 
by the bulk gravitational solution.  
By assuming a simple gravitational model, we are led to a particular $V(\phi)$ 
that is valid not just near $T_c$ 
but at all temperatures in the superfluid phase.
Moreover, from the model one can extract not only the static
thermodynamic properties, but also quantities relevant for time-dependent,
dynamic processes, like the kinetic coefficients and the correlation functions.

In the future, it would be very interesting 
to extend the numerical results above.
Two obvious directions present themselves.  
In the calculation of the speed of second sound 
from derivatives of the pressure $P$, 
one could go beyond the probe approximation 
and include the back reaction of the Abelian-Higgs sector on the metric.  
We hope that the resulting speed of second sound will be phenomenologically 
meaningful not only near $T=T_c$ but all the way down to $T=0$.  

Another interesting project for the future would be to find 
the sound wave poles in the density-density correlation function.  
By studying fluctuations of the scalar and gauge
field of the form $e^{-i \omega t + i k x}$ in the probe limit, 
one should be able to isolate
a pole of the form $1/(\omega^2 - v_2^2 k^2 + \ldots)$ 
in the Fourier transform of the retarded Green's function 
for the charge density.  Here the ellipsis denotes higher order 
(damping) terms in $k$.  Beyond the probe limit, 
there should also be a pole corresponding to the propagation of ordinary sound.

\vskip 0.2in
\noindent
{\it Note Added} --- While we were completing this paper, we learned of ref.~\cite{competitors}
which has some overlap with this work.

\vskip 0.2in
\noindent
{\it Acknowledgments} ---
We would like to thank David Huse for discussion.
P.K.K. and C.P.H. thank the INT for hospitality during the 
2008 workshop ``From Strings to Things," where this work began.
C.P.H. thanks the Galileo Galilei Institute for Theoretical Physics 
for hospitality and the INFN for partial support during the completion 
of this work.
The work of C.P.H., P.K.K., and D.T.S. was supported, in part, by the 
US NSF under Grant No.\ PHY-0756966,  NSERC of Canada,
and the US DOE under Grant No.\ DE-FG02-00ER41132, respectively.

\appendix

\section{Comparision with the Carter-Khalatnikov-Lebedev formulation
of superfluid hydrodynamics}
\label{app:CarKhal}

In this Appendix we show that the hydrodynamic equations written in
Sec.~\ref{sec:hydro} are equivalent to the set of equations proposed
previously by Israel, Carter, Khalatnikov, and Lebedev
\cite{Carter,LebedevKhalatnikov,CarterKhalatnikov1,CarterKhalatnikov2,Israel, Valle:2007xx}.
%(see also \cite{Israel, Valle:2007xx}.)
The formulations of Ref.~\cite{Carter} and
Ref.~\cite{LebedevKhalatnikov} have been shown to be equivalent in
Refs.~\cite{CarterKhalatnikov1,CarterKhalatnikov2}.  We will follow
the notation of Ref.~\cite{CarterKhalatnikov2}.

In Ref.~\cite{CarterKhalatnikov2}, superfluid hydrodynamics is
formulated as follows.  First, one postulates that the thermodynamic
properties of the superfluid are defined by a scalar function
$\Lambda$, which is a function of two 4-vectors $s^\mu$ and $n^\mu$,
which are the entropy density and the particle number density.  Since
$\Lambda$ is a Lorentz scalar, the number of variables that it depends
on is three: $s^\mu s_\mu$, $n^\mu n_\mu$, and $s^\mu n_\mu$.

One defines two Lorentz vectors $\Theta^\mu$ and $\mu^\mu$ from
\begin{equation}
  d\Lambda = \Theta_\mu ds^\mu + \mu_\mu dn^\mu\ . 
  \label{CKdLambda}
\end{equation}
The superfluid hydrodynamic equations are
\begin{eqnarray}
  & & \d_\mu s^\mu = 0 \label{CKeq1} \ ,\\
  & & \d_\mu n^\mu = 0 \label{CKeq2} \ , \\
  & & \d_\mu\mu_\nu - \d_\nu\mu_\mu = 0 \label{CKeq3} \ , \\
  & & s^\mu (\d_\mu\Theta_\nu - \d_\nu\Theta_\mu) = 0 \ . \label{CKeq4}
\end{eqnarray}

At the first sight, Eqs.\ (\ref{CKeq1})--(\ref{CKeq4}) do not bear
much resemblance to Eqs.\
(\ref{eq:Tmunu-conservation})--(\ref{eq:s-conservation}).  However,
these systems of equation are in fact equivalent.  The variables
appearing in the Carter-Khalatnikov formulation can be identified
with the hydrodynamic variables used in this paper as in the following
relations,
\begin{align}
  \Lambda &= \epsilon - f^2(\d_\mu\varphi)^2 \ , \label{CKLambda}\\
  s^\mu &= s u^\mu \ , \label{CKs}\\
  n^\mu &= n u^\mu + f^2\d^\mu\varphi \ , \label{CKn}\\
  \mu_\mu &= -\d_\mu \varphi \ , \label{CKmu}\\
  \Theta_\mu &= \frac1s \bigl[-(Ts+\mu n)u_\mu +
                 n\d_\mu\varphi\bigr]  \ . \label{CKTheta}
\end{align}
We now show that any solution to the hydrodynamic equations
(\ref{eq:Tmunu-conservation})--(\ref{eq:s-conservation}) satisfies
Eqs.~(\ref{CKdLambda})--(\ref{CKeq4}), upon the subsitutions
(\ref{CKLambda})--(\ref{CKTheta}).  First, Eqs.~(\ref{CKeq1}) and
(\ref{CKeq2}) coincide with Eqs.~(\ref{eq:s-conservation}) and
(\ref{eq:jmu-conservation}) due to Eqs.~(\ref{CKs}) and (\ref{CKn}).
Furthermore, Eq.\ (\ref{CKeq3}) is trivially satisfied by Eq.\
(\ref{CKmu}).

Now let us check Eq.~(\ref{CKdLambda}).  The left hand side is
\begin{equation}
  d\Lambda = d\epsilon - d\bigl(f^2(\d_\mu\varphi)^2\bigr) 
  = Tds + \mu dn - \d_\mu\varphi\, d\bigl( f^2(\d^\mu\varphi)\bigr),
  \label{dL1}
\end{equation}
where the thermodynamic relation $\epsilon+P = sT + \mu n$ and 
Eq.~(\ref{eq:dP}) have been used,
while the right hand side is
\begin{align}
  \Theta_\mu ds^\mu + \mu_\mu dn^\mu &= 
  \frac 1s \bigl[-(Ts+\mu n) u_\mu + n\d_\mu\varphi\bigl] (u^\mu ds + sdu^\mu)
  - \d_\mu\varphi \bigl[ u^\mu dn + ndu^\mu + d\bigl(f^2\d^\mu\varphi\bigr)
  \bigr]\nonumber\\
  &= \left(T+\frac{\mu n}s\right)ds + \frac ns u^\mu\d_\mu\varphi\,ds
  - u^\mu\d_\mu\varphi\,dn - \d_\mu\varphi\,d\bigl(f^2\d^\mu\varphi\bigr)
  \label{dL2}
\end{align}
where we have used $u^2=-1$, $u_\mu du^\mu=0$.  Now from the Josephson
equation $u^\mu\d_\mu\varphi=-\mu$ one sees immediately that
(\ref{dL2}) is identical to (\ref{dL1}).  Thus, Eq.~(\ref{CKdLambda})
is verified.

The last equation that has to be checked is Eq.~(\ref{CKeq4}).  We
first write
\begin{equation}
  s^\mu(\d_\mu\Theta_\nu-\d_\nu\Theta_\mu) = 
  \d_\mu(s^\mu\Theta_\nu) - s^\mu\d_\nu\Theta_\mu
\end{equation}
where $\d_\mu s^\mu=\d_\mu(su^\mu)=0$ has been used.
We expand
\begin{align}
  \d_\mu(s^\mu\Theta_\nu) 
  &= \d_\mu\bigl[ -(Ts+\mu n)u^\mu u_\nu + nu^\mu\d_\nu\varphi\bigr]
     \nonumber\\
  &= -\d_\mu\bigl[(Ts+\mu n)u^\mu u_\nu\bigr] + nu^\mu\d_\mu\d_\nu\varphi 
    - \d_\mu\bigl(f^2\d^\mu\varphi)\d_\nu\varphi\label{dsTheta}
\end{align}
where Eq.~(\ref{eq:jmu-conservation}) has been used;
\begin{align}
  s^\mu\d_\nu\Theta_\mu &= su^\mu \d_\nu\left[-\left(
  T+\frac{\mu n}s\right)u_\mu + \frac ns \d_\mu\varphi\right]\nonumber\\
  &= s\d_\nu T + n\d_\nu\mu + s\mu\d_\nu\left(\frac ns\right)
  + su^\mu \d_\mu\varphi \d_\nu\left(\frac ns\right) 
  + nu^\mu \d_\mu\d_\nu\varphi \ . \label{sdTheta}
\end{align}
Combining Eqs.~(\ref{dsTheta}) and (\ref{sdTheta}), using the
Josephson equation, one finds
\begin{equation}
  s^\mu(\d_\mu\Theta_\nu-\d_\nu\Theta_\mu) =
  -\d_\mu\bigl[(Ts+\mu n)u^\mu u_\nu\bigr] 
  - \d_\mu\bigl(f^2\d^\mu\varphi)\d_\nu\varphi- s \d_\nu T - n\d_\nu \mu\ . 
\end{equation}
It is easy to check that the right hand side is equal to $\d_\mu
{T^\mu}_\nu$ up to an overall sign, with ${T^\mu}_\nu$ defined in
Eq.~(\ref{eq:Tmunu-definition}), and hence is equal to zero.

%-------------------------------------------------------------

\end{document}